\documentclass[runningheads,a4paper]{llncs}
\usepackage{cite}
\usepackage{amssymb} 
\usepackage{graphicx}
\usepackage{url}
\usepackage{float}
\usepackage{soul,color}  
\usepackage{amsmath}
\usepackage{todonotes}
\usepackage{dirtytalk}
\usepackage{adjustbox}
\usepackage[inline]{trackchanges}

\begin{document}

\newcommand{\luiz}[1]{\textcolor{red}{#1}}

\title{Anomaly Detection in DevOps Toolchain}

\author{Antonio Capizzi \inst{1}, Salvatore Distefano \inst{1}, Manuel Mazzara \inst{2} \\ Luiz J.P. Ara\'{u}jo \inst{2}, Muhammad Ahmad \inst{1,3}, Evgeny Bobrov \inst{2} \\ 
\institute{University of Messina, Italy
\and Innopolis University, Innopolis, Respublika Tatarstan, Russian Federation
\and Department of Computer Engineering, Khwaja Fareed University of Engineering and Information Technology, Pakistan.
}}

\toctitle{Lecture Notes in Computer Science}
\tocauthor{Authors' Instructions}

\authorrunning{Capizzi et al.}

\maketitle

\abstract{
The tools employed in the DevOps Toolchain generates a large quantity of data that is typically ignored or inspected only in particular occasions, at most. 
However, the analysis of such data could enable the extraction of useful information about the status and evolution of the project. For example, metrics like the ``lines of code added since the last release'' or ``failures detected in the staging environment'' are good indicators for predicting potential risks in the incoming release.
In order to prevent problems appearing in later stages of production, an anomaly detection system can operate in the staging environment to compare the current incoming release with previous ones according to predefined metrics. The analysis is conducted before going into production to identify anomalies which should be addressed by human operators that address false-positive and negatives that can appear. In this paper, we describe a prototypical implementation of the aforementioned idea in the form of a ``proof of concept''. The current study effectively demonstrates the feasibility of the approach for a set of implemented functionalities.
}

\section{Introduction}

Evolution of software engineering spans over more than fifty years where different problems have been presented, and solutions explored \cite{WASSERMAN1985}. 
From  \textit{``structured programming"} to \textit{``life cycle models"} and \textit{``software development methodologies”}, researchers and developers have better understood the software development process and its complexity. Meanwhile, a fast-speed growing technological progress has transformed the usage of computers from devices for numerical and scientific computation into every-day ubiquitous devices. 
This progress has not stopped, and an increasing number of companies are moving to Agile methodologies, also including in the software development process feedback from operational stages in a DevOps \cite{Bass, DEVOPS2018} fashion.

\textit{Continuous delivery} (CD) is an important concept part of the DevOps philosophy and practice as it enables organizations to deliver new features quickly as well as to create a repeatable and  reliable process  incrementally improving to bring software from concept to customer. The goal of CD is to enable a constant flow of changes into the production via an automated software production line - the \textit{continuous delivery pipeline}. The CD pipeline has a variable complexity and can be constituted by several phases supported by different tools. However, the core idea is always the same: when a developer integrates or fixes a functionality into the software, a set of software tools automatically builds the application, starts the automatic tests and, finally, delivers the new feature. 

CD is made possible via automation to eliminate several manual routines during software production, testing, and delivery. 
CD pipeline automation 
involves in the toolchain different tools, 
each generating messages, data and logs. 
However, the amount of recorded data can prevent its manual inspection when one searches for a specific issue or traces back abnormal behavior. 
Inside a DevOps toolchain, data is generated and stored in different formats. 
The analysis of such data is a daunting task even for an experienced professional as well as its processing, recognition, mining and, consequently, addressing of critical aspects.

In this paper, we discuss how to automatically analyze the data generated during a DevOps toolchain integrated to anomaly detection (AD) methods for identifying potentially harmful software releases. As a result, software releases that can lead to potential malfunctioning during the normal system life could be identified. The implemented approach could still lead to false-positives and false-negatives since no approach can overcome this theoretical limitation \cite{books/daglib/0016921}; however, developers are provided with an instrument to validate and maintain the code. This investigation focuses on an ongoing project structured according to the DevOps philosophy, and we will apply analytical techniques to gain insights for professionals involved in the software development process.

In Section \ref{sec:bg}, background is provided, with specific regard to DevOps toolchains and AD techniques and tools.
In Section \ref{sec:anomalydetection}, we presented an approach for integrating AD into a project structured with DevOps.
After that, section \ref{sec:casestudy} describes the case study, in details: the SpaceViewer application, the corresponding DevOps process and toolchain and the developed AD module, the SpaceViewer AD system - SVADS.
Section \ref{sec:results} then reports on the experiments and obtained  results, also compared against those obtained by offline tools on the full SpaceViewer dataset, demonstrating the effectiveness of the proposed approach.
Section \ref{sec:conclusion} summarises the key aspects of the proposed approach and future work.


\section{Background}
\label{sec:bg}

This section summarizes the technical background that it is necessary in order to understand the project and its implementation. This research is bringing together two communities that not necessarily interacted much so far, therefore the respective literature and vocabulary are somehow separate, although with overlapping research. First, we discuss the details of the DevOps toolchain, then we cover those data science techniques that have been applied to the software development process. 

\subsection{The DevOps toolchain}
\label{sec:devops}
DevOps \cite{Bass} consists of a set of practices to promote collaboration between the developers, IT professionals (in particular sysadmin, i.e. who works on IT operations) and quality assurance personnel. DevOps is implemented via a set of software tools \cite{8314153} that enable the management of an environment in which software can be built, tested and released quickly, frequently, and a more reliable manner. In addition to CD, \textit{continuous integration} (CI) stands as a key concept in DevOps approaches. A typical example of CI consists of continuously integrating changes made by developers into a repository, then a project build is automatically executed, if the build works well automatic tests are started. IF also automatic tests passes, the change is integrated into the code through CD and published in production environment.

One of the main objectives of DevOps is to mitigate problems in production, which is done by reducing the gap between development and testing environments with the production environment. Collaborations between ``Dev'' and ``Ops''  aiming to reduce this gap make use of a complex toolchain including, at least, some version control tool (e.g. Git), CI/CD automation tools (e.g. Jenkins), package managers (e.g. NPM) and test tools (e.g. JUnit). Other additional tools used in DevOps are configuration management tools (e.g. Ansible), monitoring tools (e.g. Nagios), security tools (e.g. SonarCube), team collaboration tools (e.g. Jira) and database management tools (e.g. Flyway). DevOps infrastructures are typically either fully implemented on cloud platforms. It is a good practice in DevOps to build the entire infrastructure using containers; therefore, tools for containerization (e.g. Docker) are employed, sometimes coupled by tools for containers orchestration (e.g. Kubernetes).

An outcome from the complex pipeline involved in a DevOps project is the generation of a large amount of data, in particular log files and metrics generated in each stage. 
Examples of activities that generate considerable data on the project cycle include changes made by developers; the application building and its corresponding entries on the compilation and dependencies of the project; the execution of automatic tests; and software usage by end-users after release into the production.

A large amount of the data generated in a DevOps toolchain requires some form of automation and possibly dimensionality reduction and feature selection \cite{Protasov2018}.
However, collecting, storing, and analysing such a high dimensional data could enable insights into how to improve the DevOps pipeline \cite{Kontogiannis:2018}. For example, historical data can be analyzed to estimate a probabilistic measure of the success of a new release.


\subsection{Anomaly detection in software development}
\label{sec:ad-software}

The application of data science techniques to software development processes has become increasingly popular in the last decades, in part due to the availability of a growing amount of data generated during the development process. Methods like data preprocessing and machine learning have been used for tasks including estimating programming effort, predicting risks to the project and identifying defects in the produced artefacts \cite{Barakat.2019}.

In recent years, the term ``AIops'' has been coined to refer to a set of techniques which employ machine learning and artificial intelligence to enable the analysis of data from IT operations tools \cite{232959}.
As a result, there has been a noticeable improvement in service delivery, IT efficiency and superior user experience \cite{8239931}.  The literature also contains the application of AIops for DevOps to analyze data that is produced in the toolchain, specifically the operations part of DevOps \cite{aiopsdevops}. In another example, AIops has been used to support software development processes within an organization during the migration from waterfall processes to Agile/DevOps \cite{8239932}.

AD has been an increasingly popular approach for identifying observations that deviate from the expected pattern in the data. In data science, AD refers to a set of techniques used for identifying observations which occur with low frequency in the dataset, i.e. entries that do not conform to the expected distribution or pattern. Such data entries raise suspicion and represent potential risk depending on the context in which data has been collected. Examples of applications of AD in different problem domains include detection of bank frauds \cite{Guo2018}, structural defects in building construction \cite{6882174}, system health monitoring and errors in a text \cite{Chandola2016}.
It is trustworthy mentioning that there has been limited literature demonstrating the application of AD methods in the context of DevOps. An example of AD applied to DevOps operations in a Cloud platform was reported in \cite{7389388}.


\section{Integrating anomaly detection into DevOps}
\label{sec:anomalydetection}

As mentioned previously, the vast amount of data generated by the DevOps toolchain enables the use of AD techniques to reduce the probability of software errors released in production. An AD system can compare the multivariate features of the prospective release with the collected data from previous versions. The DevOps study analyzed in this work is following the development, staging, and production model. In this model, the activities are sorted in three deployment environments, detailed as follows:

\begin{itemize}
    \item \textbf{Development}: environment in which the developers work and can quickly test new features.
    \item \textbf{Staging}: testing environment to experiment and test the new features that have to be merged to the system.
    \item \textbf{Production}: environment in which the software is released and utilised by end-users.
\end{itemize}

The development and staging environments offer an opportunity for assessing the correctness of the prospective release. Moreover, the data collected during these stages enable the application of data science techniques such as AD for preventing software errors. The most suitable approach depends on the characteristics of the data.
For example, if a considerable amount of labeled data is available,  supervised learning techniques (e.g. support vector machine) can lead to satisfactory predictive accuracy. In case there is no information whether each observation in the training dataset is an anomaly, an unsupervised learning technique is the most suitable approach.

\begin{figure}[ht]
  \centering
  \includegraphics[scale=0.4]{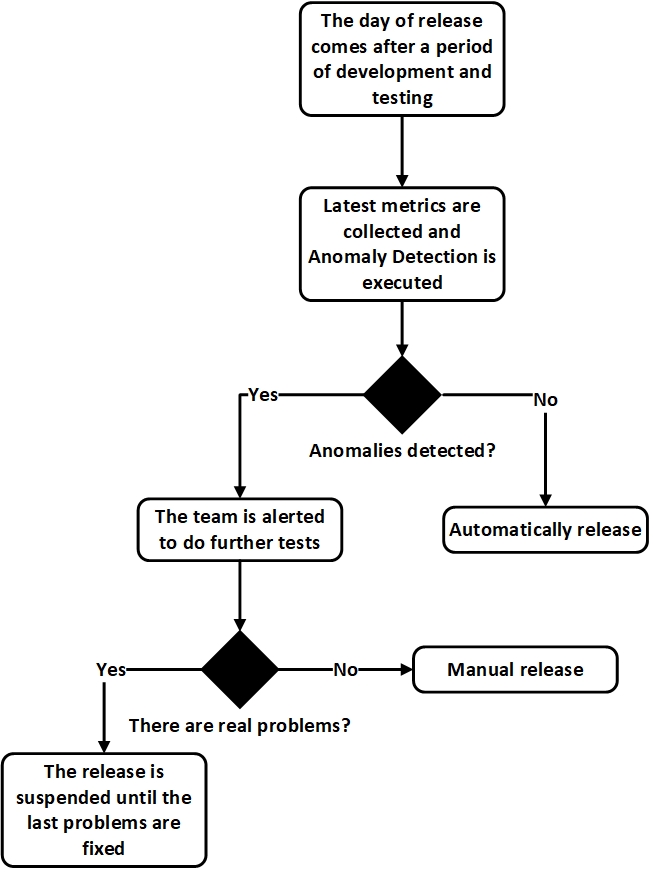}
  \caption{The anomaly detection task in the proposed DevOps workflow.}
  \label{fig:flowdiagram}
\end{figure}

This study employs the local outlier factor (LOF) algorithm, which is an unsupervised AD technique which computes the local density deviation of a multivariate data point compared to its neighbors. This method enables the identification and plotting of anomalies in the data and supports better decision-making \cite{Hodge2004}. The LOF algorithm is used before a new version of the software is moved from the staging phase to the release phase. In other words, it identifies whether the prospective release significantly deviates from exiting distributions in the following set of metrics: the number of pushes, builds and errors, lines of code that have been changed and the number of failed tests. Fig. \ref{fig:flowdiagram} shows the operational flow, which consists of three macro-phases distinguishing development, AD and recovery activities.

In the development stage, software development and testing are implemented in the development and staging environments as described previously. These activities are performed between the current release and the next version. The activities in this stage are mostly executed by the development team. In the detection stage, AD using the LOF algorithm is employed and possibly coupled with advanced computational techniques like artificial intelligence and machine learning. Moreover, the comparison of distinct AD methods can provide more a well-informed decision in the recovery phase, when a human actor assesses the identified anomalies.


\section{A case study: SpaceViewer}
\label{sec:casestudy}

\begin{figure}[ht!]
  \centering
  \includegraphics[width=0.7\textwidth]{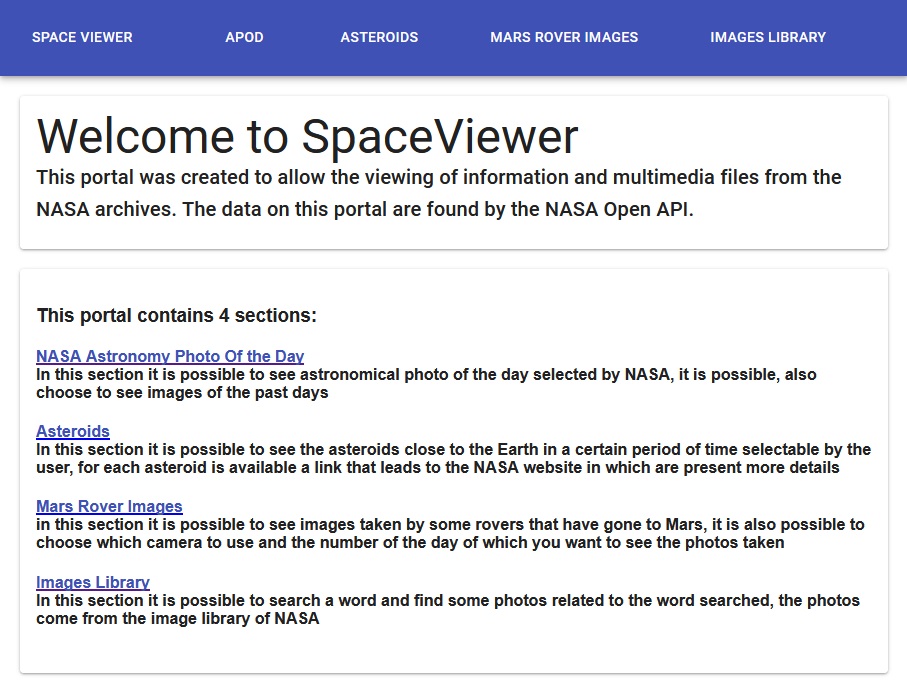}
  \caption{SpaceViewer ReactJS web-application}
  \label{fig:spaceviewerhomepage}
\end{figure}

This section describes a proof-of-concept application developed by exploiting a DevOps approach and toolchain proposed in this work. 
It consists of a Web application developed by adopting a DevOps process: \textit{SpaceViewer} \cite{spaceviewer}.
SpaceViewer is  a ReactJS \cite{Reactjs} project enabling queries for interacting and interfacing the NASA space archive exploiting their Open APIs \cite{NasaOpenAPI}. 
A client-server app has been implemented where the server-side small back-end interface \cite{spaceviewer_be} (developed in Python 3.7 \cite{Python} using Flask 1.0.2 \cite{Flask}) sends a token to the client app necessary to query the NASA DB.
Fig. \ref{fig:spaceviewerhomepage} reports the SpaceViewer homepage with the main features implemented.

\subsection{DevOps toolchain}
\label{sec:devOps-toolchain}

The  DevOps toolchain adopted in the SpaceViewer app development is composed of the following tools

\begin{itemize}
    \item \textbf{Jenkins} \cite{Jenkins}: CI/CD and automation
    \item \textbf{GitHub} \cite{GitHub}: version control
    \item \textbf{CodeClimate} \cite{CodeClimate}: assessment of the quality of the source code
    \item \textbf{Docker} \cite{Docker}: deployment tool
    \item \textbf{Slack} \cite{Slack}: team collaboration and management of automatic alerts from Jenkins Jobs
    \item \textbf{Node Package Manager - NPM} \cite{NPM}: run build, deploy, and automatic test of the ReactJS application
  \item \textbf{SpaceViewer Anomaly Detection System - SVADS} \cite{spaceviewer_ads}: this tool was created specifically for this experimentation, it will be described in the section \ref{sec:dataset}

\end{itemize}

As discussed in Section \ref{sec:anomalydetection}, the deployment environments have been implemented as follows:

\begin{itemize}
    \item \textbf{Development environment}: local in developer machines.
    \item \textbf{Staging environment}: remote server deployed in a Docker container and triggered by Jenkins. Whenever a new version of the software is pushed on the GitHub repository, the staging environment is automatically rebuilt.
    \item \textbf{Production environment}: remote server in a Docker container triggered by Jenkins. Before a build in production, SVADS is triggered.
\end{itemize}

\begin{figure}[ht]
  \centering
  \includegraphics[scale=0.70]{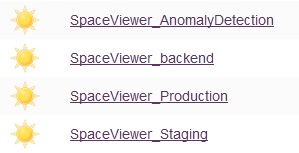}
  \caption{SpaceViewer Jenkins Jobs (Pipelines).}
  \label{fig:jenkinsjobs}
\end{figure}
The Jenkins tool has been set up to manage such deployment environments. 
Fig. \ref{fig:jenkinsjobs} depicts the Jenkins Jobs created for the SpaceViewer case study, thus establishing a  Jenkins pipelines \cite{JenkinsPipeline}.
Jenkins jobs are mainly instantiated for deploying in staging (SpaceViewer$\_$Staging) 
and production (SpaceViewer$\_$Production), while additional jobs are created to run the back-end process (SpaceViewer$\_$Backend) and  perform AD before launching the production job (SpaceViewer$\_$AnomalyDetection).
\begin{figure}[ht]
  \centering
  \includegraphics[scale=0.60]{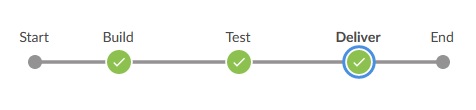}
  \caption{Staging/Production Pipelines stages}
  \label{fig:jenkinspipelinestages}
\end{figure}

The pipelines for both the Staging  and the Production deployments consist of the stages shown in Fig. \ref{fig:jenkinspipelinestages}.
An automatic system in Jenkins triggering the rebuild in Staging at every Development push on the GitHub repository has been deployed. 
As stated above, before deploying in Production, the AD job has to be performed to detect  any possible anomaly or issue in the DevOps development process.  
Then, if no anomalies are detected, the Production job is automatically triggered and the SpaceViewer software version is released in Production.
In the SpaceViewer DevOps pipeline, Jenkins is also connected through a specific plugin \cite{SlackPlugin} to the messaging software Slack \cite{Slack}. 
This way, the team can receive real-time automatic alerts regarding Jenkins jobs outcomes (e.g. failure and success). 

\begin{figure}[ht]
  \centering
  \includegraphics[width=.8\textwidth]{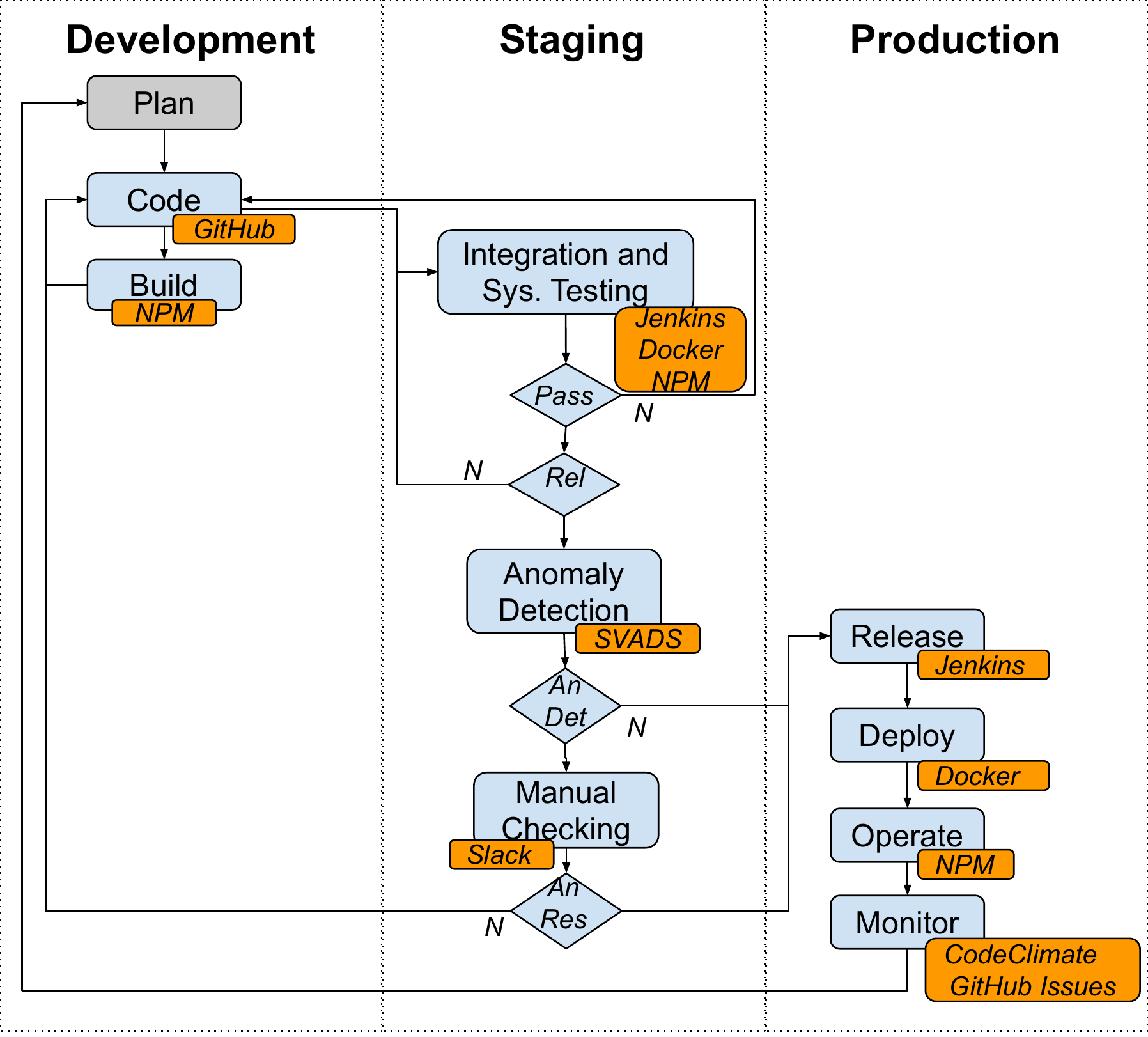}
  \caption{SpaceViewer DevOps process and toolchain.}
  \label{fig:ex-wflow}
\end{figure}

The overall SpaceViewer DevOps process and toolchain are shown in  Fig. \ref{fig:ex-wflow}, highlighting the different stages of the process and the main tools involved.  
The swim-lanes identify the three environments taken into account, correlating their activities with the different stages of the process. 
As stated above, the latter two environments are deployed into two independent containers, while the Development one runs locally into the development/developer machines.
The only step that is not directly involved in the SpaceViewer automated DevOps process is the initial Plan one.
After planning, coding activities (Code) trigger the DevOps pipeline with specific metrics from the development environment and tools (ReactJS and GitHub), as discussed in the following section.
Once implemented, SpaceViewer modules are ready for unit testing and building loop (Build exploiting the NPM tool) and, after that, they are automatically released to the Staging Environment for Integration and System Testing by the Jenkins SpaceViewer$\_$Staging job, triggered by GitHub pushes into the repository. 
This stage loops until related activities, mainly testing ones, are performed and successfully passed, then triggering the release if ready for that, always orchestrated by the SpaceViewer$\_$Staging job (see Fig. \ref{fig:jenkinsjobs}).
If so, the AD job (SpaceViewer$\_$AnomalyDetection) is launched and run the SVADS tool.
In the case of anomaly the control is demanded to the people involved in project for further Manual Checking, automatically informing the team about the anomaly through a Slack chat, the procedure of release is suspended and in production remains the latest version of application. 
On the other hand, if there are no anomalies, the  SpaceViewer$\_$Production job (see Fig. \ref{fig:jenkinsjobs}) is triggered by SVADS, the production environment is rebuilt and the latest features are integrated (Release, Deploy, Operate, Monitor) through the corresponding tools in the pipeline.

\subsection{Space Viewer Anomaly Detection System} 
\label{sec:dataset}

The tool for AD - Space Viewer Anomaly Detection System, SVADS in short - has been developed in Python and, in the SpaceViewer case study \cite{spaceviewer_ads}, consists of a script launched by the Jenkins before the delivery in Production of a new version of the software.
SVADS retrieves data relating to the last development period (i.e. since the day after the last release, to the day  the new release is being executed), generated by the DevOps toolchain and collected by the system meanwhile, 
to perform AD. 
The SVADS algorithm is mainly tasked at detecting outliers in the SpaceViewer software release to Production, to avoid potential issues for the software in Production. 
It implements the Local Outlier Factor (LOF) algorithm \cite{LOFnovelty} by exploiting the \textit{scikit-learn} Python Library \cite{scikit-learn}.
After executing the SVADS algorithm, the system fills the FLAG attribute indicating the presence/absence of an anomaly, and stores latest data in the dataset for future release AD.

Specifically, such a dataset is comprised of performance metrics collected via Rest APIs provided by the DevOps toolchain shown in Fig. \ref{fig:ex-wflow}.
The parameters taken into account by the SVADS dataset are reported below and, as discussed above, are related  to the modifications done exclusively in the last DevOps cycle:

\begin{itemize}
    \item Number of lines of code ($NLoC$) added, modified or deleted divided by the number of commits ($NCom$) from GitHub in the Code stage - $P1=NLoC/NComm$
    \item Number of builds that failed when executing the Jenkins pipeline to deploy in staging from the Integration and System Testing phase - $P2$
    \item Number of automatic tests that failed when executing the Jenkins pipeline to deploy in staging from the Integration and System Testing phase - $P3$
    \item Number of deliveries that failed when executing the Jenkins pipeline to deploy in staging from the Integration and System Testing phase - $P4$
    \item Number of issues reported by CodeClimate from the Code and Monitor phases - $P5$
    \item Number of issues reported in GitHub from Operation and Monitor phases - $P6$
\end{itemize}

Each entry in the dataset corresponds to a software release and the parameters $P1-P6$ are the number of occurrences of related events since the last release. 
They are therefore reset by any new release.
The values of such attributes are normalized according to the number of working days elapsed since the last release to mitigate the effects of longer periods of maintenance. 
It also reflects the good practice of performing regular ``small'' commits in contrast to doing few but substantial commits.
The following attributes capturing meta-data of each entry are also added to the dataset:

\begin{itemize}
    \item A unique identifier - $ID$
    \item The date of the release, i.e. when the parameter values are collected and written into the dataset - $DATE$
    \end{itemize}

Some of the above DevOps toolchain metrics are often used to also support better decision-making regarding potential risks in a software release. 
For example, a high number of failed builds, automated tests and deliveries in Staging 
might be an indicator that a specific release requires additional management effort. 
It is trustworthy mentioning that such a dataset can also enable the observation of complex patterns involving different parameters related to the occurrence of software defects, errors or faults.

It is important to point out that the SVADS tool was created for this case study, but it can be used for any project that has a DevOps Toolchain like the one used in this study.


\section{Experiments, results and discussion}
\label{sec:results}

The experimentation of the proposed approach for the DevOps toolchain in the SpaceViewer case study started in early July 2019 and took approximately one month.
In this experimentation, data entries conforming the format defined in Section \ref{sec:dataset} were added to the SpaceViewer dataset at the moment of every software release in production by the SVADS tool. 
Table \ref{tab:dataset} reports the full dataset describing 25 subsequent releases between 4\textsuperscript{th} of July and 8\textsuperscript{th} of August, uniquely identified by the attribute $ID$.

\begin{table}[ht!]
\centering
\begin{adjustbox}{width=.5\textwidth}
\begin{tabular}{|c|c|c|c|c|c|c|c|}
\hline
$P1$ & $P2$ & $P3$ & $P4$ & $P5$ & $P6$ & $ID$ & $DATE$ \\ \hline \hline 
22.57 & 0.04 & 0.06 & 0.08 & 0 & 0 & 1 & 7/4/2019 \\ \hline 
59 & 2 & 3 & 5 & 0 & 1 & 2 & 7/5/2019 \\ \hline 
87 & 1 & 4 & 6 & 0 & 1 & 3 & 7/6/2019 \\ \hline 
13 & 1 & 3 & 6 & 0 & 0 & 4 & 7/7/2019  \\ \hline 
130 & 3 & 4 & 5 & 1 & 0 & 5 & 7/8/2019 \\ \hline 
135 & 3 & 6 & 8 & 3 & 0 & 6 & 7/9/2019 \\ \hline 
27 & 2 & 4 & 7 & 6 & 0 & 7 & 7/10/2019 \\ \hline 
10 & 2 & 4 & 6 & 4 & 0 & 8 & 7/11/2019 \\ \hline 
40 & 0 & 1 & 3 & 6 & 0 & 9 & 7/12/2019 \\ \hline 
21 & 3 & 5 & 6 & 6 & 0 & 10 & 7/13/2019 \\ \hline 
33 & 3 & 5 & 6 & 6 & 0 & 11 & 7/14/2019 \\ \hline 
65 & 6 & 8 & 10 & 8 & 0 & 12 & 7/15/2019 \\ \hline 
90 & 3 & 4 & 6 & 8 & 0 & 13 & 7/16/2019 \\ \hline 
114 & 6 & 7 & 10 & 13 & 0 & 14 & 7/17/2019 \\ \hline 
255 & 5 & 9 & 9 & 12 & 0 & 15 & 7/18/2019 \\ \hline 
44 & 3 & 4 & 5 & 13 & 0 & 16 & 7/19/2019 \\ \hline 
123 & 4 & 6 & 8 & 17 & 0 & 17 & 7/22/2019 \\ \hline 
171 & 5 & 7 & 8 & 23 & 0 & 18 & 7/24/2019 \\ \hline 
100 & 3 & 4 & 5 & 23 & 0 & 19 & 7/25/2019 \\ \hline 
42 & 1 & 5 & 6 & 23 & 0 & 20 & 7/26/2019 \\ \hline 
94 & 1 & 3 & 4 & 8 & 0 & 21 & 7/29/2019 \\ \hline 
243 & 29 & 30 & 31 & 13 & 0 & 22 & 7/30/2019 \\ \hline 
28 & 5 & 6 & 8 & 15 & 0 & 23 & 7/31/2019 \\ \hline 
244 & 45 & 48 & 50 & 0 & 0 & 24 & 8/1/2019 \\ \hline 
35 & 6 & 7 & 8 & 0 & 0 & 25 & 8/8/2019 \\ \hline 
\end{tabular}
\end{adjustbox}
\caption{The SpaceViewer dataset.}
\label{tab:dataset}
\end{table}

\begin{figure}[ht]
  \centering
  \includegraphics[scale=0.65]{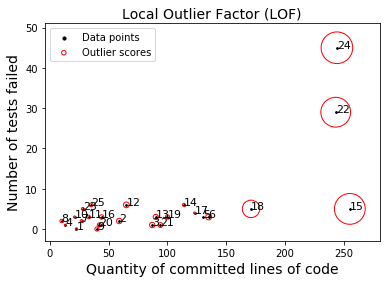}
  \caption{Outlier scores for the dataset using LOF for anomaly detection on the full SpaceViewer dataset.}
  \label{fig:lof-ad}
\end{figure}

Firstly, an initial dataset was generated to attend the requirement of a considerable quantity of observations to perform an unsupervised AD method. In this study, data concerning software releases were collected for ten days without being processed by the SVADS module.
After this initial period, the AD system was then activated, thus starting operating on the SpaceViewer DevOps process, as shown in Fig. \ref{fig:ex-wflow}. 
For each new release, the LOF algorithm was trained with the dataset comprising previous releases and the current candidate release. 
Finally, the data describing the last release is appended to the dataset and available for future use.
Fig. \ref{fig:lof-ad} illustrates the output from the LOF model after the 25\textsuperscript{th} release, i.e. the outlier scores for each observation.

Fig. \ref{fig:lof-ad} enables the observation of several insights into the integration of AD into DevOps. First, SVADS supports the identification of data entries, i.e. software releases, that clearly fails to conform expected patterns in data. For example, $ID$s 15, 22 and 24 have higher outlier scores and easily distinguished from their peers.
Second, SVADS requires some degree of human interference for labelling data with edging feature values. For example, the release with $ID$ in Fig. \ref{fig:lof-ad} is closer to most of the releases than to the clearly identified anomalies. In larger projects in the real-world, SVADS would flag such releases as requiring further assessment by the project manager.
Finally, the collection and analysis of such data enable the observation of patterns between features such as lines of codes, stages of development and occurrences of anomalies. In the implemented case study, for example, anomaly releases have been mostly identified by higher code volumes or Staging failures.

An interesting matter that deserves further consideration is whether an unsupervised AD (outlier detection) method should be employed instead of supervised AD (novelty detection). For the first case, at the moment of a new release, the AD model is trained with the entire dataset and outlier scores above a specified threshold indicate anomalies. In the second method, it is assumed that there is the availability of a significant number of software releases. Moreover, it is also necessary that each release has been labeled by a specialist (e.g. the project manager) whether it is an anomaly. Hence, the latter method can be noticed as closer to a policy-based approach for AD.


The implemented method was validated against other offline statistical and machine learning techniques. 
Several statistical methods can be utilised for identifying outliers, including the popular k-nearest neighbors and LOF.
Moreover, some AD models outperform others depending on the characteristics of the data and the problem domain. 
Fig. \ref{fig:lof-ad} illustrates four different AD models trained using the generated dataset.

\begin{figure}[ht!]
  \centering
  \includegraphics[scale=0.48]{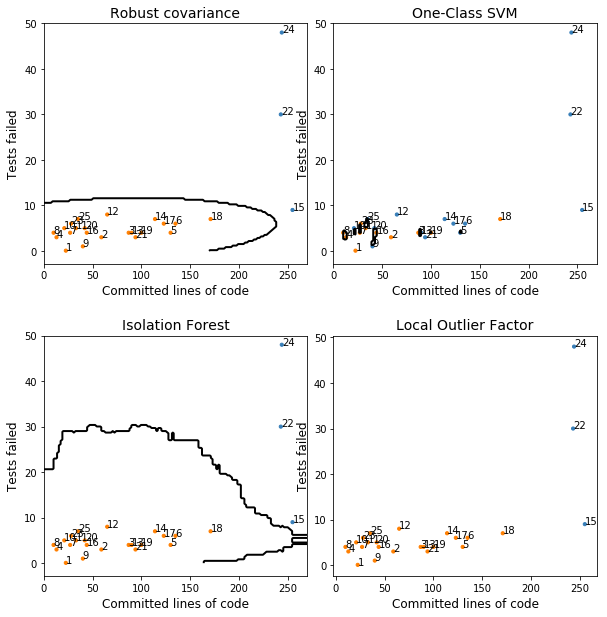}
  \caption{Comparing different AD methods and decision boundaries on the SpaceViewer Dataset.}
  \label{fig:comparing-ad-methods}
\end{figure}

These outcomes from the models in Fig. \ref{fig:comparing-ad-methods} reinforce the usefulness of the proposed SVADS approach. In fact, an ensemble of AD models enables a more precise and undisputed decision regarding software releases that are likely to result in an error in the production environment.
Finally, some AD models can provide decision boundaries for classifying anomalies which enable one to gain insights regarding which features that are more likely yo result in a risk to the ongoing project.

\section{Conclusions}
\label{sec:conclusion}

DevOps is becoming an increasingly adopted approach in software development, gaining attention from both industry and academia as  per the rising number of projects, conferences, and training programs in this field \cite{DEVOPS2018, MazzaraNSSU18}. 
A DevOps toolchain typically generates a large amount of data that enables the extraction of information regarding the status and progress of the addressing project. 
In this paper, we described a prototypical implementation of a system for detecting anomalies in software release adopting DevOps development process.

Despite the small number of functionalities implemented in our SpaceViewer case study, this paper demonstrates the feasibility of the proposed workflow. 
Obtained results and their comparison against powerful solution integrating  several AD models proves the validity of the proposed approach and its effectiveness as a tool for supporting decision-making and precise identification of potentially harmful candidate releases in the production. 
Furthermore, a dataset on AD for software release in the DevOps toolchain has been generated and made publicly available for the community.

Future work will approach the stabilization of the current implementation and broader experimentation in real-world production environments and an more extensive number of features, which has been scarcely reported in the literature. 
Moreover, future research will approach a broader discussion on how to consider the fluctuation of feature values can indicate anomalies through the project life-cycle.




\bibliographystyle{ieeetr} 
\bibliography{biblio}

\end{document}